\documentclass[aps,prl,preprint,groupedaddress]{revtex4-1}

\usepackage{graphicx}
\usepackage{dcolumn}
\usepackage{bm}
\usepackage{multirow}
\usepackage{color}%
\usepackage[english]{babel}
\usepackage{amsmath,amssymb}
\usepackage{amsfonts}
\usepackage{mathrsfs}
\usepackage{epsfig}
\usepackage{tabularx}
\usepackage{array}
\usepackage{slashed}
\usepackage{subfig}

\def\s2tw{{\rm sin ^2 \theta_{W}}}

\newcommand{\metu}{Department of Physics, Middle East Technical University, Ankara 06531, Turkey.}
\newcommand{\seven}{$70$-plet~}
\newcommand{\five}{$56$-plet~}
\begin{document}


\title{Magnetic Moments of \seven Baryons}


\author{T.M.~Aliev}
\affiliation { \metu }
\author{ V.S.~Zamiralov}
\affiliation {Skobeltsyn Institute of Nuclear Physics,Lomonosov
Moscow State University, Moscow, Russia}



\date{\today}

\begin{abstract}
Magnetic moments of the positive parity \seven baryons are
estimated within the non relativistic quark model and QCD sum rules
method. It is found that the magnetic moments of the \seven
baryons can be expressed in terms of the $D$ and $F$ couplings and
exhibit unitary symmetry. The QCD sum rules for the magnetic
moments of the \seven octet baryons are formulated. Comparison of
magnetic moments of \five and \seven baryons predicted from QCD
sum rules are presented.
\end{abstract}


\maketitle

\section{Introduction}
Study of the electromagnetic properties of hadrons represents very important source of information
about their internal structure and can provide valuable insight in understanding the mechanism of
strong interactions at low energies, i.e. about nonperturbative aspects of QCD. Particular interest
deserves magnetic moments of baryons as a subject of permanent study due to growing experimental
information~\cite{muon}.

Magnetic moments of the positive parity octet and decuplet baryons
are studied in framework of different approaches, such as
non relativistic quark model (NRQM)~\cite{muon},static quark
model~\cite{Franklin:2002xp},QCD string
approach~\cite{Kerbikov:2000sv},chiral perturbation
theory~\cite{Puglia:1999th}, skyrme model~\cite{Park:1991fb},
traditional QCD Sum rules~\cite{Chiu:1985ey},light-cone version of
QCD sum rules~\cite{Aliev:2000cy,Aliev:2002ux}, lattice
QCD~\cite{Leinweber:1992hy}.

Although the magnetic moments of the positive baryons are widely
studied experimentally and theoretically, however about their
negative parity counterparts there very limited information.
Magnetic moments of these states can be extracted through the
bremsstrahlung processes in photo and electro production reactions.

There exist comparatively few theoretical works on the magnetic
moments of the negative parity baryons in framework of
LCSR~\cite{Aliev:2014pfa,Aliev:2014foa,Aliev:2015qea}, effective
Hamiltonian approach in QCD~\cite{Narodetskii:2013cxa},
constituent quark model~\cite{Chiang:2002ah}, chiral quark model
~\cite{Dahiya:2003}, unitarized chiral perturbation
theory~\cite{Hyodo:2003dt}, chiral constituent quark
model~\cite{Sharma:2013rka} and lattice QCD~\cite{Lee:2010dq}.

The positive parity octet and decuplet baryons {\it usually} enter
the \five representation of the SU(6), which is the
group-theoretical basis of the NRQM. In the unitary SU(3)model
coupling constants of photons (and vector mesons) with baryons are
expressed in terms of the F and D constants. In chiral model
~\cite{Dahiya:2003} mixing of 56- and 70-plets are studied for
$\frac{1}{2}^+$ and $\frac{3}{2}^+$ baryons. Instead negative
parity baryons are put usually in 70-plet.

The basic element of QCD sum rules calculation is the
interpolating current of the corresponding hadron which can be
reduced to \five wave functions in non relativistic limit. Here we
note that in studying the properties of the $\frac{1}{2}^-$
baryons the interpolating current for the positive parity baryons
have been used although it has no usual NRQM limit for the 70-plet
baryons.
 Despite this difficulty we shall try to write QCD sum
rules in such a way as to respect the NRQM limit.

\section{F and D couplings in quark-diquark model}

Although the NRQM results on magnetic moments can not be expressed
in terms of F and D couplings, in~\cite{zamiralov} it was shown
that it can be achieved in framework of diquark-quark model. The
characteristic property of this model is that the photon or vector
boson field interact with diquark and single quark in a different
way.

Let us discuss the magnetic moments of \five baryons in NRQM in framework of diquark-quark model.
As an example let consider proton from \five. Its wave function can be written as
\begin{equation}
\label{first} \sqrt{18}|p, s_z=1/2> = |2u_1 u_1 d_2 - u_1 d_1 u_2
- d_1 u_1 u_2 + 2 u_1 d_2 u_1 - u_1u_2 d_1 + 2d_2 u_1 u_1 - u_2
u_1 d_1 - u_2 d_1 u_1>
\end{equation}
where subindices $1$ and $2$ means quark spin up and spin down
respectively. The wave functions of other members of octet baryons
can be obtained from proton wave function by appropriate
replacements of quark fields.

The results for the magnetic moments of the \five baryons in terms of F and D couplings can be
achieved by using their wave functions and introducing following four different matrix elements:
\begin{eqnarray}
&&<q_1 q_1,q_2^\prime |\hat{w}_q | q_1 q_1, q_2^\prime> = w^q_{11} \nonumber \\
&&<q_1 q_2,q_1^\prime |\hat{w}_q | q_1 q_2, q_1^\prime> = w^q_{12} \nonumber \\
&&<q_1 q_1,q_2^\prime |\hat{w}_q^\prime | q_1 q_1, q_2^\prime> = v^q_{11} \nonumber \\
&&<q_1 q_2,q_1^\prime |\hat{w}_q^\prime | q_1 q_2, q_1^\prime> =
v^q_{12}
\end{eqnarray}


Using these matrix elements with $e_q \hat{w}_q \sigma^q_z$ as the
modified magnetic moment operator the magnetic moments of
$\Sigma^0$ and $\Lambda$ baryons  are obtained ~\cite{zamiralov}
\begin{eqnarray}
\label{three} \mu(\Sigma^0) &&= (e_u + e_d) \frac{2}{3}w_{11} +
\frac{1}{3}e_s (2v_{11} - v_{12})
\nonumber \\
&&= (e_u + e_d) F +e_s(F-D)  \nonumber \\
\mu(\Lambda) &&= (e_u + e_d) (F-\frac{2}{3}D) + e_s(F+\frac{1}{3}D)
\end{eqnarray}
where $\frac{2}{3}w_{11} = F $,$w_{12} = D$ and $(2v_{11}-v_{12})
= 3(F-D)$ and also $v_{11}=D$, $v_{12}=3F-D$. These results show
that the non relativistic quark-diquark model reveals the unitary
symmetry pattern and the magnetic moments of baryons can be
formulated in terms of the $F$ and $D$ couplings
~\cite{zamiralov},\cite{AlievOzZam:2012rudn}.

If we put $e_u = 2/3$ and $e_d = e_s = -1/3$ from
Eq.~(\ref{three}), one can easily obtain well known $SU(3)$
relations between magnetic moments of $(1/2)^+$
baryons~\cite{Coleman:1961jn}:
\begin{eqnarray}
&&\mu(p) = \mu(\Sigma^+) = F+\frac{1}{3}D \nonumber \\
&&\mu(\Sigma^-) = \mu (\Xi^-) = -F +\frac{1}{3} D \nonumber \\
&&\mu(n) = \mu(\Xi^0) = -\frac{2}{3} D \nonumber \\
&& \mu(\Lambda) = -\frac{1}{3} D.
\end{eqnarray}

 The NRQM results can be found if we take $F = \frac{2}{3} D$, $D=1$ and replace the electric
 charge of quarks by their magnetic moments, i.e. $e_q \rightarrow \mu_q$.

Now let us analyze magnetic moment of baryons entering to the
\seven representation $SU(6)$ in framework of NRQM. The \seven in
NRQM has following decomposition $70 = (8,2) + (10,2) + (8,4) +
(1,2)$. The wave function of \seven within the NRQM obtained in
numerous works (see~\cite{Sharma:2013rka} and references therein).
Following to the~\cite{Sharma:2013rka}, the wave function of
$N^{*+}$ state in \seven with positive parity can be written as;
\begin{eqnarray}
\sqrt{18} |N^{*+}> = && |2u_1 u_1 d_2 - u_1 d_1 u_2 - d_1 u_1 u_2
+2 d_1 u_2 u_1 - u_1 u_2 d_1 -
u_1 d_2 u_1  \nonumber \\
&&+ 2 u_2 d_1 u_1 - u_2 u_1 d_1 - d_2 u_1 u_1>
\end{eqnarray}
Using this wave function, within the quark-diquark model for the
magnetic moment of $N^{*+}$  we get
\begin{eqnarray}
\mu_{N^{*+}} = &&~e_u\frac{2}{3} w_{12} + e_d \frac{1}{3}(2v_{12} - v_{11}) \nonumber \\
\end{eqnarray}

Using definitions $w_{11}, w_{12}, v_{11}, v_{12}$ (see
redefinitions after Eq.(3)) we have for magnetic moment of
$N^{*+}$
\begin{eqnarray}
\mu_{N^{*+}} &&= e_u F +e_d(2F-D)
\end{eqnarray}
Performing similar analysis for the magnetic moments of $\Sigma^{0*}$ and $\Lambda^*$ baryons
we obtain;
\begin{eqnarray}
\mu_{\Sigma^{*0}} &&= \frac{1}{2} (e_u + e_d)F +e_s(2F-D) \nonumber \\
\mu_{\Lambda^{*0}} &&= (e_u + e_d) \frac{1}{6}(9F-4D) +e_s \frac{1}{3}D
\end{eqnarray}
Transition moment $\mu_{\Sigma^{0*}\Lambda^{*0}}$ is obtained
immediately from group-theoretical relation~\cite{zamiralov,AlievOzZam:2012rudn}
\begin{eqnarray}
<\Sigma_{d\leftrightarrow s} | O | \Sigma_{d\leftrightarrow s}> -
<\Sigma_{u\leftrightarrow s} | O | \Sigma_{u\leftrightarrow s}> = \sqrt{3}<\Sigma^0| O | \Lambda>
\end{eqnarray}
where $O$ is appropriate operator and reads
\begin{eqnarray}
\mu_{\Sigma^{*0}\Lambda^{*0}} &&= -\frac{3}{2} (e_u -
e_d)(F-\frac23 D).
\end{eqnarray}
and in the NRQM yields zero.

%

Magnetic moments of other members of octet baryons can be found
with the help of appropriate replacements of quark fields. In the
limit $F= \frac{2}{3}~D$, $D=1$ and replace $e_q \rightarrow
\mu_q$ we get the NRQM results for the magnetic moments of \seven
baryons. When in Eqs. (7),(8) take the relevant quark charges we
get unitary symmetry results for the octet baryons in the \seven, similar to the relations for
the \five baryons see Eq.(4) :

\begin{eqnarray}
\mu(N^{*+}) &&= \mu(\Sigma^{*+}) = \frac{1}{3} D \nonumber \\
\mu(N^{*0}) &&= \mu(\Xi^{*0}) = F-\frac{2}{3} D \nonumber \\
\mu(\Sigma^{*-}) &&= \mu(\Xi^{*-}) = -F + \frac{1}{3} D \nonumber \\
\mu(\Sigma^{*0}) &&= -\mu(\Lambda^*) = - \frac{1}{2} (F-\frac{2}{3} D) \nonumber \\
\mu(\Sigma^* \Lambda^*) &&= -\frac{3}{2} (F-\frac{2}{3}D)
\end{eqnarray}

 Finally, magnetic moments of \seven octet baryons can be obtained from results of \five
 respectively with the help of simple replacements by comparing Eqs. (3),(7) and (8).
 For example, the magnetic moment of $\Sigma^0$
 in \seven can be
 achieved from \five one by replacing coefficients of F and D terms :\\
for $\Sigma^0$ baryons;
\begin{eqnarray}
F:\quad (e_u +e_d) && \rightarrow \frac{1}{2} (e_u +e_d +2e_s) \nonumber \\
(F-D):\quad e_s && \rightarrow e_s
\end{eqnarray}


These relations  constitute one of the main results of the present
work.

\section{QCD sum rules for magnetic moments of 56- and 70-plet baryons}

These results for the magnetic moments of octet baryons can also
be obtained from the QCD sum rules method which exhibits unitary
symmetry pattern in a sense that they can be represented in terms
of only two independent F and D type functions. Note that the
conclusion is true not only for photon but for any vector field
too (see for example~\cite{Aliev:2009ei}). The key object of the
QCD sum rules method is the interpolating currents. In the
non relativistic limit the interpolating current of baryons can be
reduced to their wave functions~\cite{Belyaev:1982sa}.

The magnetic moments of the octet baryons in framework of traditional and light cone versions of
QCD sum rules are calculated in various works (see for example~\cite{Aliev:2002ux,Chiu:1985ey}
and references therein.)


The QCD sum rules method is based upon the correlation function
$$\Pi = i \int{d^4x e^{ipx}} <0|T\{\eta_B(x) \bar{\eta}_B(0)\}|0 >_\gamma$$
where T is the time ordering operator, $\gamma$ means external electromagnetic field and
$\eta_B$ is the interpolating current carrying the same quantum numbers as the corresponding
baryon B . As an example we present the interpolating current for $\Sigma^0$ baryon

\begin{eqnarray}
\eta_{\Sigma^0} = \sqrt{\frac{1}{2}}\epsilon^{abc} \{ (u^{aT}Cs^b) \gamma_5 d^c -(s^{aT} C d^b)
\gamma_5 u^c + \beta (u^{aT}C\gamma_5 s^b) d^c - \beta(s^{aT}C\gamma_5 d^b) u^c \}
\end{eqnarray}

where $a,b,c$ are color indices, $C$ is the charge conjugation
operator and $\beta$ is the arbitrary parameter ( $\beta =-1$
corresponds to the so called Ioffe current).

In order to construct QCD sum rules for magnetic moments of the
octet baryons the correlation function is calculated in terms of
hadrons and quark-gluon degrees of freedom. By matching these two
representations the QCD sum rules for the octet baryons magnetic
moments are obtained ~\cite{Aliev:2000cy}. The magnetic moments of
octet baryons  in general form can be written as
\begin{equation}
\label{six}
\lambda^2_B \mu_B e^{-m_B^2/M^2} = e_u \Pi_1^B(u,d,s,M^2) + e_d \Pi_1^B(d,u,s,M^2) +
e_s \Pi_2 (u,d,s,M^2)
\end{equation}

where $\lambda_B$ and $\mu_B$ are the residue and magnetic moment
of corresponding baryon, respectively and $\Pi_i$ is the invariant
function in the coefficient of $\slashed{p}$$\slashed
{\epsilon}$$\slashed{q}$ Lorentz structure and their expressions
can be found in ~\cite{Chiu:1985ey,Aliev:2002ux}. Using the
$SU(2)$ symmetry Eq.(\ref{six}) can be written as
\begin{eqnarray}
\lambda^2_B \mu_B e^{-m_B^2/M^2} = (e_u + e_d) \Pi_1^B (u,d,s,M^2) + e_s \Pi_2^B (u,d,s,M^2)
\end{eqnarray}

Using the relation obtained in~\cite{Ozpineci:2003bg} which connected $\Lambda$ and $\Sigma^0$
interpolating current one can easily find the expression for the magnetic moment $\Lambda$-hyperon:
\begin{eqnarray}
3 \mu (\Lambda) = 2 \mu (\Sigma^0(d\leftrightarrow s) + 2 \mu (\Sigma^0(u\leftrightarrow s) -
\mu(\Sigma^0)
\end{eqnarray}

We can now predict the magnetic moments of the octet in \seven
with positive parity if we assume that the transformations
obtained in NRQM (see Eqs.(3),(8) and (12)) hold in the case of
QCD sum rules, i.e., at the level of correlation functions
$\Pi$'s.


In this case even when the explicit expressions for interpolating
currents of octet baryons belonging to the \seven representation
are not known one can predict the magnetic moments of these
baryons.

Using Eqs. (8),(12) as well as Eqs (15) and (16) for the magnetic
moments of $\Sigma$ and $\Lambda$ baryons in \seven we get
following sum rules


%
%
%
%
%

\begin{eqnarray}
\lambda_{\Sigma^*}^2 \mu_{\Sigma^*}^2 e^{-m_{\Sigma^*}^2/M^2} &&=
\frac{1}{2} [ e_u+e_d+2e_s ] \Pi_1^B +e_s\Pi_2^B \nonumber \\
\lambda_{\Lambda^*}^2 \mu_{\Lambda^*}^2 e^{-m_{\Lambda^*}^2/M^2} && =
[\frac{3}{2}(e_u + e_d)]\Pi_1^B + [2(e_u+e_d) - e_s]\Pi_2^B
\end{eqnarray}

 Comparing these equations with the sum rules for the $\Sigma$ and $\Lambda$ baryons from \five
 we see that they have changed drastically.

 In order to get the idea about the magnitude of the magnetic moments of \seven baryons, in the
 obtained sum rules we put their experimental mass and for residues we put their values for
 \five baryons. Under this assumption, the magnetic moments of the positive parity \seven baryons
 are given in Table 1 (the 2nd column). For completeness we also put magnetic moments of \five
 baryons in this table ( the 1st column).
\begin{table}[ht]
\begin{tabular}{|c|c|c|}
\hline
$\mu(B^*)$& QCD SR $\frac{1}{2}^+$~\cite{Aliev:2002ux} & This Work ($\frac{1}{2}^+$,\seven)   \\
\hline
$\mu(N^{*+})$& 2.72 & 0.83   \\
\hline
$\mu(N^{*0})$ & -1.65 & 0.22    \\
\hline
$\mu(\Sigma^{*+})$ & 2.52 & 0.70    \\
\hline
$\mu(\Sigma^-)$ & -1.13 & -1.13   \\
\hline
$\mu(\Sigma^0)$ & 0.70  & 0.11     \\
\hline
$\mu(\Lambda)$ & -0.50 & -0.11   \\
\hline
$\mu(\Xi^0)$ & -0.89 & 0.69   \\
\hline
$\mu(\Xi^-)$ & -1.18 & -1.18   \\
\hline
\end{tabular}
\caption{Magnetic moments of \seven and \five baryons.}
\end{table}

\section{Conclusion}
It is shown that octet baryons in the \seven can be analyzed in
the way similar to those of \five. In particular magnetic moments
are written in terms of the $D$ and $F$ quantities characteristic
for octet coupling. Moreover the main formulas for the magnetic
moments are written in such a way as to obtain the NRQM results as
well as unitary symmetry ones. Borel QCD sum rules are constructed
for the magnetic moments of the \seven octet. Comparison of
magnetic moments for 1/2$^+$ \five and \seven baryons from QCD sum
rules are presented.

\subsection{}
\subsubsection{}

\end{document}